\documentclass[referee,onecollarge,12pt]{svjour2}       
\usepackage{graphicx}
\usepackage{rotating}
\usepackage{bm}
\begin{document}
\title{
Dynamical instability causes the demise of a supercooled tetrahedral liquid
}
\author{Arvind Kumar Gautam, Nandlal Pingua, Aashish Goyal, and Pankaj A. Apte}
\institute{Department of Chemical Engineering,
         Indian Institute of Technology Kanpur,
         Kanpur,
         U.P, India  208016}
\date{21 July 2017}
\maketitle
\begin{abstract}
We investigate the relaxation mechanism of a supercooled tetrahedral liquid at its limit
of stability using isothermal isobaric ($NPT$) Monte Carlo (MC) simulations. 
In similarity with systems which are far from equilibrium but near the
onset of jamming [O'Hern et.al., Phys. Rev. Lett. {\bf 93}, 165702 (2004)], 
we find that the relaxation is characterized by two time-scales:
the decay of long-wavelength (slow) fluctuations of potential energy 
is controlled by the slope $[\partial (G/N)/\partial \phi]$ of the 
Gibbs free energy ($G$) at
a unique value of per particle potential energy $\phi = \phi_{\mbox{\tiny mid}}$.    
The short-wavelength (fast) fluctuations are controlled by the bath temperature $T$.
The relaxation of the supercooled liquid 
is initiated with a dynamical crossover after which the potential energy
fluctuations are biased towards values progressively lesser than 
$\phi_{\mbox{\tiny mid}}$.  The dynamical crossover leads to the
change of time-scale, i.e., the 
decay of long-wavelength potential energy  fluctuations 
(intermediate stage of relaxation).
Because of the condition [$\partial^2 (G/N)/\partial \phi^2 = 0$]
at $\phi = \phi_{\mbox{\tiny mid}}$,
the slope $[\partial (G/N)/\partial \phi]$ has a unique value 
and governs the intermediate stage 
of relaxation, which ends just after the crossover.
In the subsequent stage, there is a relatively rapid crystallization
due to lack of long-wavelength fluctuations and 
the instability at $\phi_{\mbox{\tiny mid}}$, 
i.e., the condition that 
$G$ decreases as configurations with potential energies lower than
$\phi_{\mbox{\tiny mid}}$ are accessed.  
The dynamical crossover point and the associated 
change in the time-scale of fluctuations 
is found to be consistent with the previous studies. 

\end{abstract}
\section{Introduction}
The end of the liquid state in the supercooled region occurs either due to glass
transition or due to the approach to the temperature corresponding to 
the limit of the stability.  In the latter case,
the mechanism of 
relaxation of the liquid (that ultimately leads to the stable crystalline state)
is generally termed as `homogeneous nucleation'.
Due to the difficulty of obtaining reproducible data at extreme conditions, there are only 
a few cases in which the relaxation point at
the stability limit has been precisely estimated.  
For example, it is well-known that 
the hard-sphere fluid exhibits jamming at a random close packed density close to
0.64, {\it independent} of the simulation or the experimental protocols.~\cite{WILLIAMS15}  
An analysis of equation of state 
points out that this density corresponds to end point of the metastable (non-crystalline) 
branch of the hard-sphere fluid.~\cite{KAMIEN07}  In the case of Stillinger--Weber~\cite{STILLINGER85} silicon,
the liquid state 
properties at the lower limit of stability of 1060 K~\cite{LUEDTKE88,LUEDTKE89} have been
precisely estimated based on free energy computations.~\cite{APTE12,APTE15}

Here we attempt to study the 
relaxation at the stability limit
of supercooled water modeled by
mW potential of Molinero and Moore.~\cite{MOLINERO09}
mW is a coarse-grained model of water in which 
water is represented by a single atom with 3-body (tetrahedral)
interactions to mimic the effect of hydrogen bonds
of real water. 
mW potential  is a
modified form of Stillinger-Weber potential in which the strength
of the 3-body term is higher than in the case of Silicon.  Due to its
short-ranged interactions, simulations of water using mW model are
much faster than the simulations
using other commonly used atomistic models of water.~\cite{MOORE09}
Moreover, mW model correctly  reproduces the following properties:~\cite{MOORE09}
 the stability of hexagonal ice and its melting point, the enthalpy
of melting, the liquid density at the melting point and at 298 K, the
maximum density of the liquid, and the liquid-vapor surface tension.
However due to lack of hydrogen atoms, the rate of crystallization
of supercooled
mW water is higher compared to that of 
atomistic models of water.~\cite{MOORE09}  Although mW predicts unrealistic
crystallization rates, due to its computational efficiency, 
it is used to study relaxation in 
the supercooled water to gain qualitative insights into the behavior of 
real water. Recently, mW model was
used to analyze the transitions between high density and low density amorphous
ices in supercooled
water.~\cite{LIMMER14}

The lower limit of stability 
of the supercooled mW water has been
estimated to be close to 200 K at a pressure of 1 bar.~\cite{LIMMER11}  
It is important to note that 
independent studies yield the same average density of 0.98 gm/cm$^3$ at the stability 
limit,~\cite{LIMMER11,HUJO11} below which it is not possible to equilibrate the liquid.  
Isobaric cooling simulations using Molecular Dynamics
with rates varying over an order of magnitude (from 1 to 10 K/ns), yield the same limiting value
of enthalpy (-43.03 kJ/mol)~\cite{MOORE11-2} at 202 K.  Moreover, the rate of relaxation as measured by
the rate of decrease of enthalpy, density and the rate of increase in 4-coordinated particles is found
to be the maximum at the stability limit.~\cite{HUJO11}  
Inspite of a large number of studies on mW liquid, it is
not clear if nucleation is the relevant mechanism of 
relaxation at the limit of stability.  
In an attempt to identify the critical nuclei (at 205 K and 1 atm pressure), 
Moore and Molinero found that the identified nuclei had a broad range of 
shapes and crystallization probabilities.  Hence it was 
concluded that ``other reaction coordinates, such as the 
structure of the liquid wetting the nuclei,
are relevant to define the transition states of ice crystallization'' [Please see caption
of Supplementary Figure 7 of Ref.~\cite{MOORE11-2}].  
In effect, critical nuclei for ice formation  could not be successfully identified,
which suggests that nucleation is not the relevant mechanism by which relaxation is initiated
at or near the limit of stability, $T_s \approx 205 K$.   

In this work,  we find that relaxation mechanism 
differs in one crucial aspect from the nucleation phenomena.
In the latter, the unstable state [consisting of the metastable (mother) phase with 
a critical nucleus] is an {\it equilibrium} state, i.e., it corresponds 
to a {stationary} condition of the free energy.  In contrast to this, 
we show that the instability that leads to relaxation of the liquid occurs 
across a unique {\it dynamical} (i.e., non-stationary) point of the free energy function.
Further, the relaxation mechanism is found to have a close similarity to
sheared foam systems, which are far from equilibrium but near the onset of jamming.~\cite{ONO02,OHERN04}
In such systems, there are two-time scales: the decay of long-wavelength fluctuations (relaxation)
is controlled by the `effective' temperature while the fluctuations
over the shorter time scale are controlled by the bath temperature.~\cite{OHERN04}  
We find that in the supercooled mW liquid, the relaxation is initiated 
with a change of time-scales
or the dynamical crossover, which signals the decay of long-wavelength fluctuations.  
The latter is controlled by the unique configurational temperature 
or equivalently, the
slope $[\partial (G/N)/\partial \phi]$ of the Gibbs free energy function $G(T,P,N,\phi)$.
The properties (the per particle potential energy, the density,
and the fraction of the 4-coordinated particles)
at the crossover point are found to be unique (independent of the system size) and 
agree with the earlier studies.~\cite{MOORE09,HUJO11,MOORE11-2,LIMMER11}

\section{Methodology}

Following the method of the earlier work on SW-Si,~\cite{LUEDTKE89,APTE15} 
we studied the relaxation mechanism of supercooled mW liquid
at 205 K (which is a more accurate estimate of the 
limit of stability~\cite{MOORE11-2}) 
and zero pressure in isothermal isobaric ($NPT$)
Monte Carlo (MC) simulations.  Each MC simulation step, on average, 
consisted of N particle displacement attempts and 2 volume change attempts.
A simulation box of cubic shape, and with periodic boundary conditions 
across all faces was used.
The properties of the largest network of 4-coordinated particles 
along the trajectories were also computed (in a manner similar to Ref.~[5] and [6]).  
To trace the network, we considered two
particles as bonded, if the distance between these particles 
is 1.4 $\sigma$ or less.  

The total potential energy $\Phi$ of the mW particles is 
computed as follows:~\cite{STILLINGER85,MOLINERO09}
\begin {equation}
\Phi(r_{1},....,r_{N}) = \sum_{ij} v_2 + \sum_{ijk} v_3,
\label{eq:phitotal}
\end{equation}
where $v_2$ and $v_3$ are the two-body and three-body potential energy terms respectively.  
The per particle potential energy is computed as $\phi = \Phi/N$, where $N$ is the total number
of particles.   
The term $\sum_{ij}$ represents sum over all pairs of particles and the term $\sum_{ijk}$ 
represents sum over all triplets of the particles.  The expression for the two-body and 
three-body terms are given by the following expressions
\begin{eqnarray}
v_2(r_{ij}) = \epsilon f_2({r_{ij}}/{\sigma}) \\
v_3(r_{i},r_{j},r_{k}) = \epsilon f_3({r_{i}}/{\sigma},{r_{j}}/{\sigma},{r_{k}}/{\sigma}) ,
\label{eq:v2v3}
\end{eqnarray}
where $r_i$,
$r_j$, and
$r_k$ are the position vectors of the particles, and $\epsilon$ and $\sigma$ are the energy and 
length parameters, respectively. The functions $f_{2}$ and $f_{3}$ are given by the following 
expressions:~\cite{STILLINGER85}
\begin{equation}
f_2(r) = \left\{ \begin{array}{ll}
                 A(B r^{-p} - r^{-q})\exp[(r-a)^{-1}] & ,\mbox{if $r<a$} \\
                 $0$ & \mbox{otherwise} 
                 \end{array} 
            \right. 
\label{eq:ff2r}
\end{equation}
and
\begin{equation}
f_3(r_{i},r_{j},r_{k}) = h(r_{ij},r_{ik},\theta_{jik})+h(r_{ij},r_{jk},\theta_{ijk})
                         +h(r_{ik},r_{jk},\theta_{ikj}) 
\label{eq:f3r}
\end{equation}
where $r_{ij}$ and $r_{ik}$ etc. are the distances between the particles and $\theta_{jik}$ is 
the angle between vectors ${\bf r_{ij}}$ and ${\bf r_{ik}}$.  
The expression for $h$ contained in the 3-body term $f_3$ is given by
\begin{equation}
h(r_{ij},r_{ik},\theta_{jik}) = \lambda\exp[\gamma (r_{ij}-a)^{-1} + \gamma (r_{ik}-a)^{-1}]
                                (\cos\theta_{jik}+1/3)^2
\label{eq:hr}
\end{equation}
The values of the parameter used in the above equations are $A=7.049556277$, $B=0.6022405584$, $p=4$, $q=0$,
$a=1.8$, $\lambda=23.15$ and $\gamma=1.20$.~\cite{STILLINGER85} The values of the energy and the length 
parameters in the mW potential are as follows: $\epsilon = 6.189$ kcal/mol, $\sigma=0.23925$ nm.~\cite{MOLINERO09} 
Throughout this
work, all the quantities (unless stated otherwise)
are expressed in dimensionless units in terms of mW potential
parameters $\sigma$ and $\epsilon$.~\cite{MOLINERO09}

The 3-body energy 
of a given triplet of particles ($ijk$) consists of 3-terms
corresponding to the  angles $\theta_{jik}$,
$\theta_{kji}$, $\theta_{ikj}$
centered at the particles $i$, $j$, and 
$k$ [see Eqs.~(\ref{eq:f3r}) and (\ref{eq:hr})].  
The three body energy was assigned to a particle of
the triplet as follows: the particle at which the 
bond angle is centered is
assigned the 3-body energy resulting from the particular bond-angle, for 
example, the term 
$h(r_{ij},r_{ik},\theta_{jik})$ was assigned to particle $i$ of the triplet
.~\cite{LUEDTKE89}
The total 3-body energy assigned to a particle 
$(\phi_i^{3B})$
consisted of contributions from all triplets involving that particle.
The per particle 3-body energy of the largest 4-coordinated network of a given instantaneous
configuration
was computed as $\phi_{4c}^{3B} = \left(\sum_{i=1}^{N_{4c}} \phi_i^{3B}\right)/N_{4c}$, where
$\phi_i^{3B}$ is the 3-body energy of particle $i$ in the network, and
$N_{4c}$ is the total number of particles in the network.  
 
\hspace{2ex}
The main emphasis in this work is on the potential energy distributions. 
We compute the logarithm of the potential energy 
distribution generated by a trajectory as follows:
$\log p(\phi) + \mbox{constant} = \log N_c (\phi)$, where $N_c$ is the number
of configurations sampled in the entire NPT-MC trajectory with the per particle 
potential energy in the range
    $(\phi-\Delta \phi/2)$ and 
    $(\phi+\Delta \phi/2)$; 
   $\Delta \phi = 3 \times 10^{-4}$ is the
   width of the bin.   The symbol `$\log$' represents the natural logarithm.
In order to analyze the various stages of relaxation of the liquid, 
we compute the intermediate distributions at certain points along the trajectories.
The intermediate distributions (denoted here by $'$)are given by 
$\log N_c' (\phi)$, where $N_c'(\phi)$ is the number
is the number of configurations at a given $\phi$ sampled
upto a particular point along the trajectory.  Please note that $N_c' (\phi) \le N_c (\phi)$, where
$N_c (\phi)$ is the total number of configurations at a given $\phi$ sampled along the entire 
trajectory.  To differentiate it from intermediate distributions, we use
the term `final' distribution for  $p(\phi)$.   Throughout this work $p(\phi)$ is obtained
from a single (sufficiently long) trajectory only for a given system size $N$.  
The distributions are not averaged over
independently generated trajectories.  
The reason for this has been explained in detail 
in the last (`Summary')  section.
When analyzing the potential energy distributions, we focus on the straight line regions (SLRs)
that appear in the intermediate distributions.  To locate a SLR, we consider the intermediate
distribution that yields the linear correlation coefficient ($R^2$)
of the straight line fit 
that is closest to unity ($R^2 \rightarrow 1$). As we shall see in the next section, the
SLRs are not just the geometric features of the distributions but these have a physical meaning
in that the trajectory shows irreversible changes (or relaxation) across the mid-point of
each of the SLRs.

\section{Relaxation of the supercooled liquid}

\hspace{2ex}
We generated, by trial and error, NPT-MC trajectories in which
relaxation is delayed the most.  We have generated 3 such
trajectories at 205 K with system sizes of N=10648, 4096, and 1000 particles.
For completeness, we also include data from the shorter 
trajectories with 10648 and 4096 particles.
The data-sets and the figures resulting from the 5 trajectories 
are as follows:

Data set (1): Figures 1--5 show the data generated by the longer
trajectory with $N=10648$ particles.  

Data set (2): Figures S1--S6 (supplementary information file) 
are generated using the data from the longer trajectory with $N=4096$ particles.  

Data set (3): Figures S7--S12 (supplementary information file) 
are generated using the data from the trajectory with $N=1000$ particles.  

Data set (4): Figures 6--8 are generated using the data from 
a shorter trajectory with $N=10648$ particles.  

Data set (5): Figures S13--S17 (supplementary information file) 
are generated using the data from a shorter
trajectory with $N=4096$ particles.  

\hspace{2ex}
In this section, we describe the relaxation process 
observed in the longer trajectories [Data sets (1)--(3)].  
The relaxation processes during the shorter 
trajectories [Data sets (4) and (5)] are described in a separate section.  
Figure~\ref{fig:t205N10648_full} 
shows the block averages of the per particle potential energy ($\phi$), 
the fraction of the particles in the largest 4-coordinated network ($f_{4c}$),
and the fraction of the 4-coordinated particles ($f_4$) along the NPT-MC
trajectory with 10648 particles [Data set(1) listed above].  
Figure~\ref{fig:rho205N10648_full} shows the block averages of the density ($\rho$)
and the per particle 3-body energy of the largest 4-coordinated 
network  ($\phi_{4c}^{3B}$) along  the same trajectory.
The block averages are taken over $10^5$ MC steps.  
Figure~\ref{fig:lfit205N10648_old} shows the evolution of the potential energy
distributions generated by the trajectory. 
The point along the trajectory at which the logarithm of the potential energy
distributions develops the straight line region (SLR) 
(see Figure~\ref{fig:lfit205N10648_old}) 
is marked as `SLR' point in Figs.~\ref{fig:t205N10648_full} and ~\ref{fig:rho205N10648_full}.
The red symbols in Fig.~\ref{fig:t205N10648_full} represent the microstates
with the per particle potential energy in the interval
$[
     \phi_{\mbox{\tiny mid}}
       ]=(
      \phi_{\mbox{\tiny mid}}-\Delta \phi/2,
      \phi_{\mbox{\tiny mid}}+\Delta \phi/2
     )$, 
where
$\phi_{\mbox{\tiny mid}}$
is the mid-point of the SLR as indicated in Fig.~\ref{fig:lfit205N10648_old}.
In this work, we point out that the relaxation of the liquid 
is a two stages process as described below. 
 
\flushleft Stage (i) The intermediate (short-time) relaxation is 
governed by the slope $[\partial (G/N)/\partial \phi]$ 
of the Gibbs free energy function $G(T,P,N,\phi)$ with a unique
value at $\phi_{\mbox{\tiny mid}} = -1.760 \pm 0.001$.    
The value of the slope is unique because of the condition 
$[\partial^2 (G/N)/\partial \phi^2] = 0$ at  
$\phi = \phi_{\mbox{\tiny mid}}$, as we shall see later.
This stage of relaxation 
ends just after the dynamical crossover of potential
energy fluctuations across $\phi_{\mbox{\tiny mid}}$, 
which results in the change
in the time-scale of fluctuations.  

\vspace{1ex}

\flushleft Stage (ii) The relaxation in this stage is relatively rapid due to
lack of long-wavelength fluctuations, and due to instability at 
$\phi_{\mbox{\tiny mid}}$, i.e., 
the condition that $G$ decreases as the microstates
with potential energy lower than
$\phi_{\mbox{\tiny mid}}$ are accessed.
The relaxation occurs over a relatively longer time
and ultimately results in the formation of
the stable crystalline phase.

\vspace{2ex}

\hspace{2ex}
In this work, our main focus is on the intermediate [stage (i)] relaxation.
The trajectory is propagated from an initial configuration which is highly
disordered ($\phi > \phi_{\mbox{\tiny mid}}$), and which is far from equilibrium.  
In the initial part of the trajectory, the potential energy and density
fluctuate around the local minimum [$\phi_m, \rho_m$] of the probability distribution.
At a certain stage (just before the R point as clearly seen 
in Figs.~1, 2, S1, and S2),  there is a  
increase in the local tetrahedral order (i.e., a decrease in $\phi_{4c}^{3B}$) 
and in the size of the 4-coordinated network (measured by $f_{4c}$).  
Simultaneously, the potential energy fluctuations
approach $\phi_{\mbox{\tiny mid}}$.
The local minimum [$\phi_m, \rho_m$] is 
statistically less accessible in this region.  
After the dynamical crossover point (dashed vertical line, 
see Figs.~1, S1, S7), 
the potential energy fluctuations are 
biased towards values progressively lesser
than $\phi_{\mbox{\tiny mid}}$.  Simultaneously,
there is a change of time scale of fluctuations 
which is associated with
a second order discontinuity in the Gibbs free energy 
function (as we shall see later).  The 
intermediate relaxation stage ends just after the
crossover (at the SLR point in Figs.~1, S1, and S7)
and results in the formation of a straight line region (SLR) 
in the potential energy distributions 
at $\phi_{\mbox{\tiny mid}}$ (see Figs.~3, S3, and S9).

\hspace{2ex}
To demonstrate the dynamical crossover 
more clearly, we have shown a zoomed
in portion of the trajectory in Fig.~\ref{fig:t205N10648}.  Here the block averages 
are taken over a smaller number of MC steps (13000 MC steps) as compared to that in 
Fig.~\ref{fig:t205N10648_full}.  Before the dashed vertical line, the potential 
energy fluctuations are biased towards  values slightly greater 
than  $\phi_{\mbox{\tiny mid}}$.  
After the crossover  (i.e., the dashed vertical line), 
the fluctuations are biased towards values progressively lesser 
than $\phi_{\mbox{\tiny mid}}$.  
The changes in the block averages across
$\phi_{\mbox{\tiny mid}}$
are seen to be irreversible.  
Since the governing feature of the 
free energy is [$\partial (G/N)/\partial \phi$] with a configurational
temperature not equal to bath temperature ($T_c \ne T$), 
the irreversible decrease of potential energy
suggests that relatively fast (or short wavelength) 
fluctuations (responsible for equilibration with respect to 
bath temperature $T$) survive at the crossover.
Thus there is a change in the time-scale of fluctuations, i.e., dissipation of 
long-wavelength potential energy fluctuations at the crossover. The relatively 
rapid and irreversible decrease in the block average potential energy (see Figs.~1, S1, and S7) 
after the crossover
supports our proposition about the decay of long-wavelength fluctuations.

\hspace{2ex}
The observation that the slope 
$[\partial (G/N)/\partial \phi]$ (or equivalently, $T_c$)
at $\phi_{\mbox{\tiny mid}}$
governs the overall
evolution of the system (intermediate relaxation)
while the bath temperature controls the
short-wavelength fluctuations seems to be very similar to the concept of `effective' temperature
in non-equilibrium systems close to jamming.~\cite{OHERN04,ONO02}  In their study, Ono et. al. 
plotted the probability distribution $\log_{10} \Omega(U)$ of
configurational energies $U/N$ obtained by numerical simulations of 
sheared foam system (see Fig. 4(a) of Ref.~\cite{ONO02}).  The `effective' temperature
was obtained from the slope of the straight line region (SLR)
of the distribution (see Figs.~4a and 4b in Ref.~\cite{ONO02}).
In such systems,
the 
fluctuations that decay over a longer time (which results in
intermediate relaxation in our work) are related
to the effective temperature~\cite{OHERN04} 
(which is similar to the $T_c$ obtained
from $[\partial (G/N)/\partial \phi]$ in our case),
while the faster fluctuations are related to the bath temperature ($T$).

\hspace{2ex}
As for the structural changes, we find that 
the fluctuations in the size of the 4-coordinated network diminish and
there is a irreversible increase in the average size of the network
(measured by  increase in ${\langle f_{4c} \rangle_b}$)
across the crossover.  Simultaneously there is an irreversible increase 
in the local tetrahedral order 
(i.e., decrease in ${\langle \phi_{4c}^{3B} \rangle_b}$ below a value of $\approx 0.22$)
as seen in  Figs. 2 and 5.  
Overall, it seems that the crossover is caused by 
the cooperative structural 
changes involving 4-coordinated particles (i.e., development of 
a certain threshold of the local tetrahedral order).
A similar dynamical crossover with accompanying changes in the local tetrahedral order
is also observed in case of trajectories with 
$N=4096$ (see Figs.~S1, S2, S5, S6) and $N = 1000$ (see Figs.~S7, S8, S11, S12).  
The values of per particle potential energy, the density (see 
Figs.~\ref{fig:rho205N10648}, S6, and S12), and the network properties at
the crossover point are obtained by linear interpolation at the dashed 
vertical lines in the figures. These values are reported in Table I.   
These properties seem to be independent of the system size $N$.   

\hspace{2ex}
The dynamical crossover results in the end of the liquid state.
This conclusion is supported by the fact that the local minimum 
{${[\phi_m, \rho_m]}$} of the probability distribution
is no longer accessible in a statistically
significant manner after the crossover, i.e., the microstates
with potential energy and density in the range {${[\phi_m, \rho_m]}$}
are negligibly smaller in number after the crossover as compared
to the number of such microstates before the crossover.  
As can be seen in Fig.~\ref{fig:t205N10648_full} and ~\ref{fig:t205N10648}, there
is not a single microstates corresponding to
{${[\phi_m, \rho_m]}$}
accessible after the
crossover.  Same is the case for trajectory with $N = 4096$ particles (see
Figs.~S1 and S5).  
In case of the trajectory with $N = 1000$ particles, there is just a single microstate
with potential energy and density in the range
{${[\phi_m, \rho_m]}$} (see Figs.~S7 and S11).  
The lack of long-wavelength potential energy fluctuations at the 
crossover is responsible for the inaccessibility
of the the local minimum {${[\phi_m, \rho_m]}$}.  Thus the 
`R' point in all 3 trajectories is located very close to the location of the crossover.

\hspace{2ex}
Now, we point out that the dynamical crossover and the associated change in the 
time-scale of fluctuations is consistent with earlier studies.  In these studies,
the instability was detected in isobaric MD simulations in which the liquid was 
supercooled at a certain rate.  
Most importantly, it was found that the rate of changes
of potential energy, density, and the fraction of 4-coordinated particles 
is the {\it maximum} at the instability limit.  
The maximum in the rate of changes is consistent
with our conclusion about the dissipation of large-scale fluctuations.  
Moore and Molinero~\cite{MOORE11-2} have reported a maximum
in the rate of decrease of enthalpy at 1 atm pressure at 202 K during isothermal
isobaric (NPT) Molecular Dynamics (MD) 
cooling simulations at rate of 1 K/ns.  This maximum occurs
at a per particle potential energy value of $-1.759$ in reduced units. 
This agrees with the dynamical crossover point found in our work
$\phi_{\mbox{\tiny mid}} = -1.760 \pm 0.001$.
The value of -1.759 in reduced units (in the work by Moore and Molinero) 
is computed by using the enthalpy
value of -43.03 kJ/mol at 202 K and 1 atm
(as read at the location of the dashed vertical line in Fig.~1a of Ref.~\cite{MOORE11-2})
and a density of 0.45 (in reduced units). 
In the study of Limmer and Chandler,~\cite{LIMMER11}  the density was
found to be 0.98 gm/cm$^3$ ($\approx 0.449$ in reduced units) at the limit of
stability of the mW liquid.  Further, the rate of change of density with temperature was
found to be the maximum at the stability limit
(see the density plot on the extreme left in the lower panel of Fig.~2 in Ref.~\cite{LIMMER11}).
Hujo et. al.~\cite{HUJO11} have reported 
a sharp decrease in density  across a value
of $0.45$ ($\sigma^3$) in MD cooling simulations (see the temperature-density curve with
parameter value of 23.15 in Fig.~1 of Ref.~\cite{HUJO11}), 
which agrees with the value at the dynamical crossover (Table I).  Further, Hujo et. al.
have mentioned that the maximum in the rate of change of density 
($d\rho/dT)$, the fraction of 4-coordinated
particles ($d f_4/dT$), the tetrahedral order parameters ($d Q/dT$), and
$(d H/dT)$ (as measured by the maximum in $C_p$) coincide at the limit of
stability  (as stated at the end of Section 3.3 and in Section 
3.4 in Ref.~\cite{HUJO11}).
Moore and Molinero have reported that 
at 0.98 gm/cm$^3$ (i.e. $0.449 \sigma^3$)
a maximum in the rate of change of density 
with respect to temperature occurs (Fig.~1 of Ref.~\cite{MOORE09}). 
In the same study, a maximum in the rate of 
increase of 4-coordinated particles in MD cooling simulations ($d f_4/dT$)
is found to be at $f_4 \approx 0.75$
for a cut-off radius of $3.35 ^o A$ (i.e., 1.4 $\sigma$) (this value was read from 
the lower panel of Fig.~3 in Ref.~\cite{MOORE09}). This is close to 
the average value of $\langle f_{4} \rangle \approx 0.742$ reported in Table I.   
Moore and Molinero have also noted that maxima in the rate of increase 
of tetrahedrality order parameter ($d Q/dT$), 
fraction of 4-coordinated particles ($d f_4/dT$),  density $(d\rho/dT)$ 
and the maximum structural correlation length $\xi$ occur 
at the same temperature $T_{LL} = 202 \pm 2$ K.~\cite{MOORE09} The coincidence
of the maxima in $d f_4/dT$, $d Q/dT$ and $d H/dT$ in the MD cooling simulations suggests that
cooperative changes in the 4-coordinated particles cause these maxima.  This agrees
with our observation that the change in the time-scale of potential energy fluctuations
(at the dynamical crossover)
is accompanied by a relatively rapid and irreversible changes
in $\langle \phi_{4c}^{3B} \rangle_b$,
$\langle f_{4c} \rangle_b$, and $\langle f_4 \rangle_b$.

\hspace{2ex}
The size of the block averages $\langle \phi \rangle_b$
is crucial in detecting the crossover and  the structural properties 
at the crossover (see Table I).  There is a irreversible decrease in 
$\langle \phi \rangle_b$ across $\phi_{\mbox{\tiny mid}}$ 
during the crossover, as seen in zoomed-in portion of 
the longer trajectory at each $N$ (Figs.~4, S5, S11).  If block averages are taken over a 
larger number of MC steps (for e.g., $10^5$ MC steps  as in the full trajectory Fig. 1), 
the crossover cannot be detected precisely.  
On the other hand, if the block averages are taken over very small number of MC steps,
due to influence of short-wavelength fluctuations (which are responsible for equilibration
with respect to $T$), the block averages do not exhibit
an irreversible decrease at the crossover.  In such cases, the properties at the
crossover cannot be determined unambiguously.    
Thus there is a minimum and maximum limit on the size of the block averages in relation to the
dynamical crossover.
In our study, the size of the block averages 
in the zoomed-in
portions of the trajectories is decided (by trial and error) 
so that the irreversible change in 
$\langle \phi \rangle_b$
at the crossover is as steep as possible.  

\hspace{2ex}
It is important to note that while the properties in Table I are independent 
of the system size $N$,
the length of the trajectory upto `R' point or equivalently the lifetime 
of the liquid is strongly
dependent on $N$.  The length of the trajectory upto `R' point is 
much shorter for $N=10468$ (Fig.~1) particles
than that for $N=1000$ (Fig.~S7) particles.  This can be rationalized as follows.
As the system size $N$ increases, 
the potential energy distribution becomes
progressively steeper. 
As such, when starting from a higher potential energy (disordered) configuration
(with $\phi > \phi_m$), the
approach to equilibrium (and hence the dissipation of large-scale potential energy fluctuations) 
is expected to be faster, resulting in the shorter trajectories with increase in $N$.
This is consistent with what we observe.   On the other hand, the
properties at the dynamical crossover itself (listed in Table I) are independent of the system
size.  This is expected because 
$\phi_{\mbox{\tiny mid}}$ (per particle potential energy at the crossover)
as well as the slope of the Gibbs free energy 
$\partial G/\partial \Phi = \partial (G/N)/\partial \phi$ 
at the crossover are both intensive properties.

\hspace{2ex}
The evolution of the system after the first stage of the relaxation 
(i.e., after the SLR point in Figs.~1, S1, and S7) constitutes the  
2nd stage of relaxation. 
Our data shows that after the SLR point, 
the configurations with potential
energies greater than $\phi_{\mbox{\tiny mid}}$ are not accessed in a 
statistically significant manner, i.e., the logarithms of SLR distribution
and the `final' distribution closely match for potential energies 
greater than 
$\phi_{\mbox{\tiny mid}}$ 
(see Figs.~\ref{fig:lfit205N10648_old}, S3, and S9) for all system sizes.  
This shows that the long-wavelength fluctuations are completely dissipated at
the end of the intermediate (first) stage of relaxation.  The relatively rapid
changes in the second stage (after the SLR point, see Figs.~1, S1 and S7) 
is caused by 
the lack of long-wavelength fluctuations and the instability
at  $\phi_{\mbox{\tiny mid}}$, i.e., the condition that Gibbs free energy
$G$ decreases as configurations with potential energies lower than
$\phi_{\mbox{\tiny mid}}$ are accessed.  
If the trajectories are continued for sufficiently long
time, the system would ultimately convert into the 
stable crystalline state.
Moore and Molinero~\cite{MOORE11-2} have shown that crystallization times 
(i.e., time required to crystallize about 70 \% of the liquid sample)
are the minimum at or around the limit of stability (see Time-temperature transformation 
diagram, Fig.~2a of Ref.~\cite{MOORE11-2}).
In order to explore the structural changes responsible for the minimum crystallization times, a 
study of crystallization in the 2nd stage of relaxation using global
order parameters such as $Q_6$~\cite{STEINHARDT83}
will be interesting and will be pursued in a future work.

\section{Gibbs free energy as a function of $\phi$}

\hspace{2ex}
The logarithm of the probability distribution $\log p(\phi)$
can be equated to the
Gibbs free energy provided that the maximum possible microstates 
at a given $\phi$ are sampled by the trajectory.  
Thus such a distribution should be unique.
That our generated distribution is unique,
at least for  $\phi \ge \phi_{\mbox{\tiny mid}}$,
is supported by two facts :  \\ 
\vspace{1ex}
(i) The average per particle potential energies and average densities 
of the liquid (i.e., cumulative
averages upto R point in Figs.~1, 2, S1, S2, S7, and S8) are found to be: 
$\langle \phi \rangle$ = -1.7542 ($N$=10648), -1.7547 ($N$=4096), -1.7539 ($N$=1000), and
$\langle \rho \rangle$ = 0.4495 ($N$=10648), 0.4494 ($N$=4096), 0.4495 ($N$=1000).
These values are fairly independent of $N$.
The liquid distribution (i.e., intermediate distribution upto R point) and the final 
distribution are the same  for $\phi > \phi_{\mbox{\tiny m}}$  (see Figs.~3, S3, and S9). 
Thus, the agreements of the average properties for different $N$ 
points to the uniqueness of the final potential energy distributions, at least, 
for  potential energy values greater than $\phi_{\mbox{\tiny m}}$. 
\\
\vspace{1ex}

(ii) The mid-point of the SLR 
$\phi_{\mbox{\tiny mid}}$ 
is -1.7592, -1.7614, and -1.7590 for 
trajectories with $N=$10648, 4096,
and 1000 particles, respectively.  
The value of configurational temperatures
$T_c$ of the SLR are 204.17 K, 203.48 K, and 203.69 K for
trajectories with $N=$10648, 4096,
and 1000 particles, respectively  
(see Figs.~3, S3, and S8).  Thus, both 
$\phi_{\mbox{\tiny mid}}$ and $T_c$ are independent of $N$.

\hspace{2ex}
Since the `final' distribution
and the SLR distribution are the same for $\phi \ge \phi_{\mbox{\tiny mid}}$, the above
two observations indicate the uniqueness of the distribution at least for potential
energy values greater than $\phi_{\mbox{\tiny mid}}$. 
Thus we conclude that 
$G(T,P,N,\phi) = - k_B T \log p(\phi) + \mbox{constant}$ for 
$\phi \ge \phi_{\mbox{\tiny mid}}$, where $\log p(\phi)$ 
refers to the `final' probability distributions in Figs.~3, S3, and S9.  
The inflection point (at or just above $\phi_{\mbox{\tiny mid}}$)
seen in the probability distributions 
(Figs.~3 and S3 for N=10648 and 4096 trajectories)
is therefore, also the  
inflection point of the Gibbs free energy.

\hspace{2ex}
We find that there is a discontinuity
in the second order derivative of $G(T,P,N,\phi)$ 
at the mid-point of the SLR.
This can be seen by the fact that the potential energy distribution can be curve-fitted
by the Taylor series expansion around the mid-point of the SLR in the following form:
\begin{eqnarray}
   \log~p_{\mbox{\tiny SLR}} (\phi) &=& \log~p_{\mbox{\tiny SLR}} (\phi_{\mbox{\tiny mid}}) 
               +a_1(\phi-\phi_{\mbox{\tiny mid}})
               +\frac{1}{3!}a_3(\phi-\phi_{\mbox{\tiny mid}})^3 \\ \nonumber
               &+&\frac{1}{4!}a_4(\phi-\phi_{\mbox{\tiny mid}})^4 
               +\frac{1}{5!}a_5(\phi-\phi_{\mbox{\tiny mid}})^5,
\label{eq:prob_density1}
\end{eqnarray}
where
\begin{equation}
       a_i = \left.
              \frac{\partial^i \log p_{\mbox{\tiny SLR}}(\phi)}{\partial \phi^i}
             \right|_{\phi =\phi_{\mbox{\tiny mid}}}.
\label{eq:prob_density2}
\end{equation}
Here $p_{\mbox{\tiny SLR}}(\phi)$ is the intermediate distribution upto the
SLR point.
Note that in the above expansion we have taken the second order 
term to be zero, i.e., $a_2 = 0$, 
since this condition is satisfied at the mid-point of the SLR.
In Fig.~\ref{fig:lfit205N10648_old}, we have fitted the SLR distribution with 
$a_3 = 5 \times 10^{-5}$ and $a_4 = -6 \times 10^{-6} $.   
The curve fit is reasonably good for potential energies less than
the mid-point of the SLR 
($\phi < \phi_{\mbox{\tiny mid}}$).   
In the case of trajectories with system sizes of 4096 and
1000 particles, multiple SLRs appear simultaneously (see Figs.~S3 and S9).  
A Taylor series expansion can be fitted to each of these SLRs as shown 
in Figs.~S4 and S9.  
The Taylor series expansion around 
$\phi=\phi_{\mbox{\tiny mid}}$, on the one hand,
fits the distribution at lower potential energies 
$\phi \le \phi_{\mbox{\tiny mid}}$ (at least
upto the next SLR at the lower potential energies), 
while on the other hand, it deviates from the actual distribution
at the higher potential energies $\phi > \phi_{\mbox{\tiny mid}}$.
Since $\log p_{\mbox{\tiny SLR}}(\phi) = \log p(\phi)$ for 
$\phi \ge \phi_{\mbox{\tiny mid}}$,
the Taylor series expansion shows a discontinuity in
the Gibbs free energy function $G(T,P,N,\phi)$
at $\phi_{\mbox{\tiny mid}}$.   
The straight line 
fit to the SLR (with $\phi_{\mbox{\tiny mid}} \sim -1.759$) shows that the 
Gibbs free energy function has a continuous first order derivative.
This enables us to conclude that there 
is a discontinuity 
in the second order derivative 
of $G(T,P,N,\phi)$ with respect to $\phi$  at 
$\phi_{\mbox{\tiny mid}}$.  

\hspace{2ex}
We now discuss the important issue of weather the SLR implies that
the second
order derivative of Gibbs free energy $[\partial^2 (G/N)/\partial \phi^2]$ 
is analytically zero at $\phi_{\mbox{\tiny mid}}$.  
That this is the case
is supported by the following observations: 
(i) the Taylor series expansion fits well for 
$\phi \le \phi_{\mbox{\tiny mid}}$ by considering the second order derivative to be zero 
[$a_2 = 0$ in Eq.~(7)],
(ii) the presence of inflection point 
[in case of $N=10648$ (Fig.~3) and $4096$ (Fig.~S3) particle trajectories] 
means a change in the sign of the curvature
around  $\phi_{\mbox{\tiny mid}}$, which implies 
that the second order derivative is zero, and 
(iii) the first order derivative $[\partial (G/N)/\partial \phi]$
at $\phi_{\mbox{\tiny mid}}$ is independent of $N$ (i.e., 
$T_c \approx 203.8 \pm 0.5$ K has a unique value as seen in Figs.~3, S3, and S9),
which also suggests that it is an extremum value, i.e.,
the second order derivative is zero.
Based on these observations, we conclude that 
analytically the second order derivative 
is zero [i.e., $\partial^2 (G/N)/\partial \phi^2 = 0$]
at $\phi_{\mbox{\tiny mid}} = -1.760 \pm 0.001$.  

\hspace{2ex}
Now we comment on the SLRs found at potential energies lower than
$\phi_{\mbox{\tiny mid}} = -1.760 \pm 0.001$ in case of $N=4096$ and
$N=1000$ trajectories (see Figs.~S3 and S9).
We observe that the trajectories exhibit irreversible changes across
the mid-point of each of those SLRs : in the case of 4096 particle trajectory,
there is an irreversible decrease in the block average potential energy 
across the mid-point  
$\phi_{\mbox{\tiny mid}}' = -1.768$ (Fig. S3) of the second SLR as seen in Fig. S5;
while in the case of 1000 particle trajectory there is an irreversible decrease across
the mid-point 
$\phi_{\mbox{\tiny mid}}''' = -1.7731$ of the fourth SLR (Fig.~S8) as seen in Fig. S10.
These irreversible changes occur just after the SLR point as seen in Figs.~S5 and S10
and highlight the fact (as also mentioned in the methodology section) that SLRs
correspond to physical changes along the trajectory.
If any of such  SLRs 
results from the analytical condition $[\partial^2 (G/N)/\partial \phi^2] = 0$,
the slope $[\partial (G/N)/\partial \phi]$ of the SLR
should be independent of $N$, 
since both $(G/N)$  and $\phi$ are intensive properties.  
Thus such SLRs should be seen in `final' distributions of {\it all}
system sizes.
However, the very fact that the SLRs at the lower potential energy
are only observed for small system sizes  
shows that  these SLRs do not correspond to the
analytical feature of the free energy function.  
Hence lower potential energy SLRs are 
not the inflection points of $G(T,P,N,\phi)$. 
Such SLRs appear due to broadening of the distribution with smaller size, and 
are transient features, i.e., such SLRs do not survive in the `final' 
distributions. \\

\section{Relaxation in the shorter trajectories}

\hspace{2ex}
Now we consider the results from the shorter trajectories
with 10648 (Figs.~6--8)   and 4096  (Figs.~S13--S17) particles.
The 10648 particle trajectory in terms of block averages over $50000$ steps is shown
in Fig.~6.  
The SLR is formed with a mid-point
at a slightly higher value
$(\phi_{\mbox{\tiny mid}}^{\mbox{\tiny{s}}} > -1.760)$ 
(see Fig.~7) than in the case of the longer
trajectory.  
The changes in the density and the per-particle 3-body energy along the
trajectory are shown in Fig.~8.
There are two aspects to consider  :(i) Does the  relaxation
in the shorter trajectory occur due to the same physical phenomena ? 
(ii) As noted in Section 4 the probability distribution 
is `unique' for $\phi > \phi_{\mbox{\tiny mid}}$ in the case of the
longer trajectory.  The crucial question is
what is the reason behind  the uniqueness of the distribution ?  We attempt to
answer these questions  by analyzing the shorter trajectories.

With regard to point (i), we find that shorter trajectory shows the same relaxation
behavior as in the case of the longer trajectory: 
complete dissipation of fluctuations occurs 
after the dynamical crossover across the value of $\approx -1.760 \pm 0.001$.
In Figs.~6 and 8, the dashed vertical line is drawn at a location 
(dynamical crossover)  beyond which
the potential energy fluctuations are biased towards energies progressively 
lesser than the unique value of $-1.760 \pm 0.001$. 
Figure~6 shows that there is a  large irreversible decrease in the potential energy 
after the crossover (dashed vertical line).  
This implies that large scale fluctuations of potential energy are 
completely dissipated (similar to the case of the longer trajectory),
after the 
dashed vertical line.   The values of 
$\langle \rho \rangle_b$ (see Fig.~8), 
$\langle f_{4c} \rangle_b$, 
and $\langle f_4 \rangle_b$ (see Fig.~6) at the
crossover are listed in row 4 of Table I.  
These values are also consistent with those
from the longer  trajectory (row 1 of Table I).
Qualitatively similar results are obtained in the case of 
the shorter trajectory with 4096 
particles (Figs.~S13--S17 and row 5 of Table I). 

To address  the point (ii) above, 
we observe that in the shorter trajectory
initial relaxation occurs 
across 
$\phi_{\mbox{\tiny mid}}^{\mbox{\tiny{s}}}$: 
before the SLR point, 
the block averages
in Fig.~6 are greater than 
$\phi_{\mbox{\tiny mid}}^{\mbox{\tiny{s}}}$; 
while after the
SLR point the block averages
in Fig.~6 are lesser than 
$\phi_{\mbox{\tiny mid}}^{\mbox{\tiny{s}}}$.
Similarly, the block averages of density and per particle 3-body 
energy (Fig. 8) show irreversible changes after the SLR point.
The irreversible changes indicate that after the SLR point
the Gibbs free energy ($G$)
of the configurations with 
$\phi < \phi_{\mbox{\tiny mid}}^{\mbox{\tiny{s}}}$  is lower as
compared to the value of $G$ of the configurations with
$\phi > \phi_{\mbox{\tiny mid}}^{\mbox{\tiny{s}}}$.
As a result of the relaxation across
$\phi_{\mbox{\tiny mid}}^{\mbox{\tiny{s}}}$,
the liquid region of the trajectory (i.e., the region where microstates corresponding
to the local minimum of the probability distribution [$\phi_m^s, \rho_m^s$] are
accessible) is dominated by configurations with
$\phi > \phi_{\mbox{\tiny mid}}^{\mbox{\tiny{s}}}$.
In contrast to this, the liquid region in the longer trajectory (Fig.~1)
corresponds to configurations with 
$\phi > \phi_{\mbox{\tiny mid}}$.
Thus, the longer trajectory samples
all possible configurations  with 
$\phi > \phi_{\mbox{\tiny mid}}$ yielding `unique' liquid properties.

\section{Summary}
\hspace{5ex}
In this work, we examined the relaxation of supercooled mW liquid 
at a temperature (205 K) corresponding to the limit of stability
at zero pressure.  
Starting with different initial configurations, we generated 
sufficiently long  NPT-MC trajectories
with system sizes of $N$ = 10648, 4096, and 1000 particles.  We find that
the relaxation of the liquid occurs in two stages: (i) the intermediate relaxation
which is governed by the unique value of $[\partial (G/N)/\partial \phi]$
at  
$\phi_{\mbox{\tiny mid}} = -1.760 \pm 0.001$ and (ii) the relatively rapid 
relaxation which
is triggered by the instability at $\phi_{\mbox{\tiny mid}}$, i.e., by the 
condition that $G$ decreases as the
configurations with potential energies lower than 
$\phi_{\mbox{\tiny mid}}$ are accessed.  
In this work, we focused on the intermediate [stage (i)] relaxation which ends just
after the dynamical crossover, i.e., at the SLR point along the trajectory
(see Figs.~1, S1, and S7).  
The crossover results in the end of the liquid state, i.e.,
the local minimum of the Gibbs free energy is not accessible after the crossover
due to decay of long-wavelength fluctuations.   
After the crossover, there is an irreversible increase in
the size of the 4-coordinated network 
(measured by increase in $\langle f_{4c} \rangle_b$) 
and the local tetrahedral order of the network 
(measured by decrease in $\langle \phi_{4c}^{3B} \rangle_b$).
These changes suggest that the relaxation is associated with cooperative structural 
changes involving the 4-coordinated particles.  The dynamical crossover 
and the corresponding change in the time-scale of fluctuations
is consistent with previous isobaric MD cooling simulations~\cite{MOORE09,HUJO11,MOORE11-2,LIMMER11}.  
We also find that the dynamical crossover is associated with a discontinuity
in the second 
order derivative of Gibbs free energy [$\partial^2 (G/N)/\partial \phi^2$] 
at $\phi_{\mbox{\tiny mid}} \approx -1.760 \pm 0.001$.

\hspace{2ex}
The relaxation mechanism is 
qualitatively similar to that
found recently for SW-Si liquid at its limit of stability 1060 K and 
zero pressure.~\cite{APTE15} Further the threshold value of per particle 3-body
energy of the network at the crossover $\langle \phi_{4c}^{3B} \rangle_b \approx 0.22$
(see Figs.~5, S6, and S12)
is comparable to the value of $0.21$ 
found in case of SW-Si at the instability point.~\cite{APTE15}
This suggests that the dynamical crossover (i.e., the change in the time-scale) of
the fluctuations occurs due to the system acquiring a certain threshold of the 
local tetrahedral environment.  As in the case of SW-Si,~\cite{APTE15} 
the liquid state at the limit of stability is not a metastable equilibrium state.  
Rather, it can be viewed as 
a {\it constrained} equilibrium state, where
the constraint is imposed by the time required to develop the threshold 
local tetrahedral environment which 
results in the dynamical crossover.  
The unique liquid properties are obtained using the {\it optimal} trajectory in which 
the approach to the crossover is delayed the most, which allows the system
to explore all possible microstates
with potential energies greater than
$\phi_{\mbox{\tiny mid}} = -1.760 \pm 0.001$
{\it before} the SLR point.  The longer trajectory for each $N$
is a {\it near-optimal} trajectory that closely satisfies the above condition.
This is clear from the observation that the `final' distribution
and the `SLR' distribution coincide 
for $\phi > \phi_{\mbox{\tiny mid}}$ 
in the case of longer trajectory at each $N$ (see Figs.~3, S3, and S9).
In the case of the shorter  (sub-optimal) trajectory, 
because of the initial crossover
at $\phi_{\mbox{\tiny mid}}^{s}$ (see Figs.~6, S13, and S16) 
all possible microstates
with $\phi > \phi_{\mbox{\tiny mid}} $ cannot be sampled.

\hspace{2ex}
As mentioned in the Methodology section, the potential energy distribution 
$p(\phi)$ in our work has been obtained from a single (near-optimal) trajectory only 
for each $N$; $p(\phi)$ is not averaged over several independent trajectories.
When one generates a set of 
trajectories starting from different initial configurations,
only one trajectory approaches  the optimal trajectory
(in the sense described above)
most closely, and the others are sub-optimal.  
Hence, averaging the distribution from all the trajectories is not appropriate.  
This is consistent with what  
Limmer and Chandler have mentioned: ``For conditions of liquid
instability, (i.e., the no-man's land here there is no barrier between
liquid and crystal), the method of rare-event sampling is no longer appropriate.
$\cdots$ 
The results depend upon the initial preparation of the system because 
the unstable system is far from equilibrium" 
(please see page 134503-8 of Ref.~\cite{LIMMER11})
As an extension of the present work, it will be interesting to explore the
structural origin of the threshold local tetrahedral environment 
that leads to the rapid crystallization
after the dynamical crossover.

\begin{acknowledgements}
P.A. is grateful to Professor B. D. Kulkarni for encouragement
and guidance on the phenomenology of the dynamical instability.
This work was supported by the young scientist scheme of the 
Department of Science and Technology, India.  \\
\end{acknowledgements}

%
%
\clearpage
\begin{center}
\begin{table}
\caption{\label{tab:delg1}
The dynamical crossover point at 205 K and zero pressure for system sizes of N=10648, 4096, and 1000 particles. 
The third column reports the values of mid-point of the SLRs as seen in Figs.~3, S3, S9, 7, and S15.  
In columns 4--7,
the values at the dynamical crossover point are reported.  The 
values are obtained by linear interpolation of the block averages across 
the dashed vertical lines in the corresponding figures.
The last two rows, marked by (*), correspond to the shorter trajectories.
    } 
\vspace{5ex}
\begin{tabular}{|c|c|c|c|c|c|c|}
  \hline
{$T(K)$} & {$N$} 
& {$\phi_{\mbox{\tiny mid}}$} 
& {${\langle \phi \rangle}_{b}$} 
& {${\langle \rho \rangle}_{b}$} 
& {${\langle f_{4c} \rangle}_{b}$}  
& {${\langle f_{4} \rangle}_{b}$}   \\ \hline \hline
205 & {10648} & {-1.7592} & -1.7590 & 0.4489 &   {0.729} &  0.741 \\ \hline \hline      
205 &  {4096} & {-1.7614} & -1.7608  & 0.4491  & {0.731} &  0.743 \\ \hline \hline
205 &  {1000} & {-1.7590} & -1.7591 &  0.4489 &  {0.731} &  0.743 \\ \hline \hline
205 & {10648*} & {-1.7562} & -1.7599 & 0.4489 &   {0.726} &  0.739 \\ \hline \hline      
205 &  {4096*} & {-1.7564} & -1.7614  & 0.4491  & {0.731} &  0.744 \\ \hline \hline
\end{tabular}      
\end{table}
\end{center}
%
\clearpage
\begin{figure}
  \caption{\label{fig:t205N10648_full}
  The NPT-MC trajectory at 205 K (with $N = 10648$ particles) and zero pressure.
  The block averages $\langle \cdots \rangle_b$ are over $10^5$ MC steps.  Here $\phi$ is the
  per particle potential energy (in reduced units), and $f_{4c} = N_{4c}/N$ (please refer to the
  right ordinate) is the fraction 
  of particles in the largest connected network of 4-coordinated particles.
   The blue symbols denotes the points along the trajectory at which the local 
   minimum $(\phi_m, \rho_m)$ 
   of the probability distribution (corresponding to the liquid state) is accessed.  
   This minimum
   is located within the rectangular area formed by the points 
   $(\phi - \Delta \phi/2, \rho- \Delta \rho/2)$ and $(\phi+\Delta \phi/2, \rho+\Delta \rho/2)$, 
     where
   $\phi = -1.75465$, $\rho = 0.44934$, $\Delta \phi = 4.4 \times 10^{-4}$, and 
   $\Delta \rho = 1.2 \times 10^{-4}$.   
   The point along the trajectory at which straight line region is formed in the
   potential energy distribution (see Fig.~\ref{fig:lfit205N10648_old}) is denoted as `SLR'. 
   The red symbols denotes the point along the trajectory at which the per particle
   potential energy ($\phi$)
   corresponds to the mid-point of the SLR, i.e., to the interval
   $[
      \phi_{\mbox{\tiny mid}}
       ]=(
      \phi_{\mbox{\tiny mid}}-\Delta \phi /2,
      \phi_{\mbox{\tiny mid}}+\Delta \phi /2
     )$.
   The vertical dashed line corresponds to the dynamical crossover point 
   along the trajectory as explained in the text.
   }
\end{figure}
\begin{figure}
  \caption{\label{fig:rho205N10648_full}
  The block averages of the density $\rho$, and 
   the per particle 3-body energy of largest 4-coordinated
  network ${\phi_{4c}^{3B}}$
  are shown for the same trajectory as in 
   Fig.~\ref{fig:t205N10648_full}.  
   The   other   symbols have the same meaning as in 
   Fig.~\ref{fig:t205N10648_full}.  
   }
\end{figure}
\begin{figure}
  \caption{\label{fig:lfit205N10648_old}
   The potential energy distributions generated by the trajectory in 
   Fig.~\ref{fig:t205N10648_full}.  
   The blue stars represents intermediate distribution 
   upto `R' point (see Fig.~\ref{fig:t205N10648_full}).  
   The intermediate distribution (denoted by pink square symbols)
    containing the straight line region (black squares) 
    is the distribution upto `SLR' point
   (see Fig.~\ref{fig:t205N10648_full}).  
   The `x' (green) symbols represents {\it final} distribution obtained from trajectory upto 
   5 million MC steps.  
   By {\it final}, we mean that the distribution is not expected to evolve 
   further in  the range of $\phi$ values in
   the above figure due to irreversible changes leading to crystallization.   
    The values of the correlation
    coefficient $R^2$ of the straight line fit to the SLR region (black squares),  
    the mid-point $\phi_{\mbox{\tiny mid}}$,
      and the configurational (or effective) temperature  $T_c$
      of the SLR region (black squares)
    are given in the inset.   
   }
\end{figure}
\begin{figure}
  \caption{\label{fig:t205N10648}
  The zoomed in portion of the trajectory in Fig.~\ref{fig:t205N10648_full}.  Here 
  $\langle \cdots \rangle_b$ denotes block averages over $13000$ MC steps.  The other symbols have 
   the same meaning as in Fig.~\ref{fig:t205N10648_full}
   }
\end{figure}
\begin{figure}
  \caption{\label{fig:rho205N10648}
  The zoomed in portion of the trajectory in Fig.~\ref{fig:rho205N10648_full}.  Here 
  $\langle \cdots \rangle_b$ denotes block averages over $13000$ MC steps.  The other symbols have 
   the same meaning as in Fig.~\ref{fig:rho205N10648_full}
   }
\end{figure}
\begin{figure}
  \caption{\label{fig:t205N10648_short}
  The shorter NPT-MC trajectory with $N = 10648$ particles.
  The block averages $\langle \cdots \rangle_b$ are over $50000$ MC steps.  
   The local minimum of the probability distribution $[\phi_m^{s}, \rho_m^{s}]$ 
   is located within the rectangular area formed by the points 
   $(\phi_m^{s}-\Delta \phi/2, \rho_m^{s}-\Delta \rho/2)$, 
   and 
   $(\phi_m^{s}+\Delta \phi/2, \rho_m^{s}+\Delta \rho/2)$, 
   where
   $\phi_m^{s} = -1.7531$, $\rho_m^{s} = 0.44976$, $\Delta \phi = 4.4 \times 10^{-4}$, and 
   $\Delta \rho = 1.2 \times 10^{-4}$.   
   The dashed vertical line corresponds to the dynamical crossover point 
   along the trajectory beyond which the potential energy fluctuations are biased
   towards energies lower than $-1.760 \pm 0.001$.  The red symbols represent the microstates
   with potential energy in the interval  
   $[ 
    \phi_{\mbox{\tiny mid}}^{\mbox{\tiny s}}
       ]=(
    \phi_{\mbox{\tiny mid}}^{\mbox{\tiny s}}
      -\Delta \phi/2,
    \phi_{\mbox{\tiny mid}}^{\mbox{\tiny s}}
      +\Delta \phi/2
     )$, where
    $\phi_{\mbox{\tiny mid}}^{\mbox{\tiny s}}$ is the mid-point 
   of the SLR (see Fig.~\ref{fig:lfit205N10648_short}) that appears
   in the intermediate  distribution upto the SLR point.
   }
\end{figure}
\begin{figure}
  \caption{\label{fig:lfit205N10648_short}
   The potential energy distributions generated by the trajectory in 
   Fig.~\ref{fig:t205N10648_short}.   
   All symbols have the same meaning as in Fig.~\ref{fig:lfit205N10648_old}.
   Note that the mid-point of the SLR does not correspond to the unique value obtained
   from the longer trajectory.   
   The {\it final} distribution (`x', green symbols) is obtained from trajectory upto 
   3.2 million MC steps.  By {\it final}, we mean that the distribution is unlikely to evolve 
   further in the range of $\phi$ values shown in the above figure.   
   }
\end{figure}
\begin{figure}
  \caption{\label{fig:rho205N10648_short}
  The block averages of the density $\rho$, and 
   the per particle 3-body energy of largest 4-coordinated
  network ${\phi_{4c}^{3B}}$
  are shown for the same trajectory as in 
   Fig.~\ref{fig:t205N10648_short}.   The block averages are taken over 50000 MC steps. 
   The  symbols have the same meaning as in 
   Fig.~\ref{fig:rho205N10648_full}.  
   }
\end{figure}
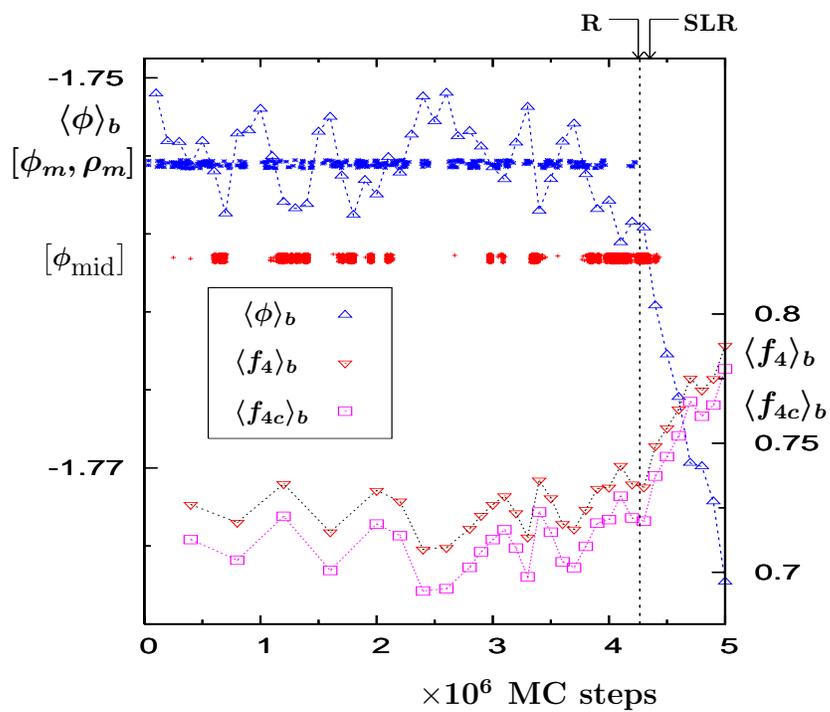
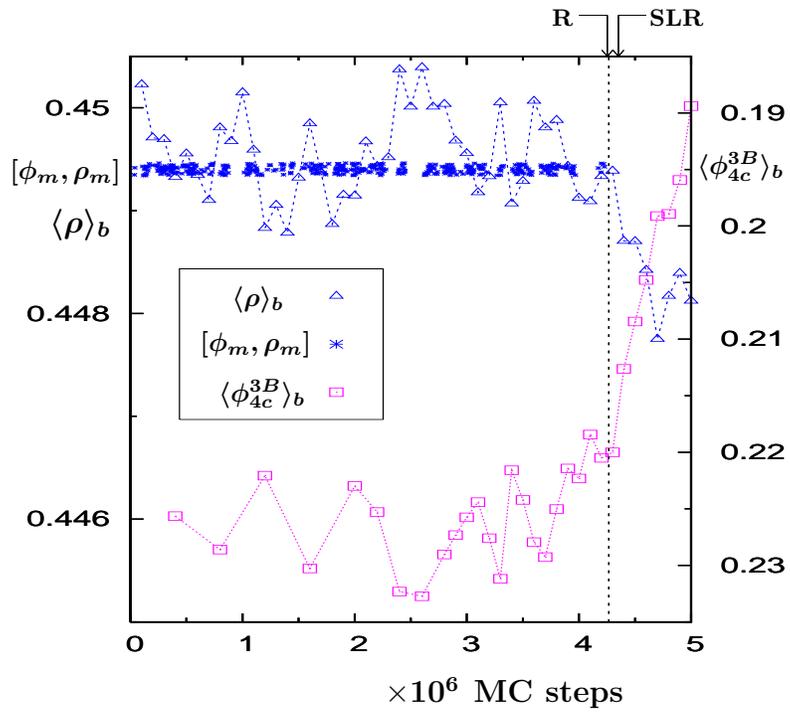
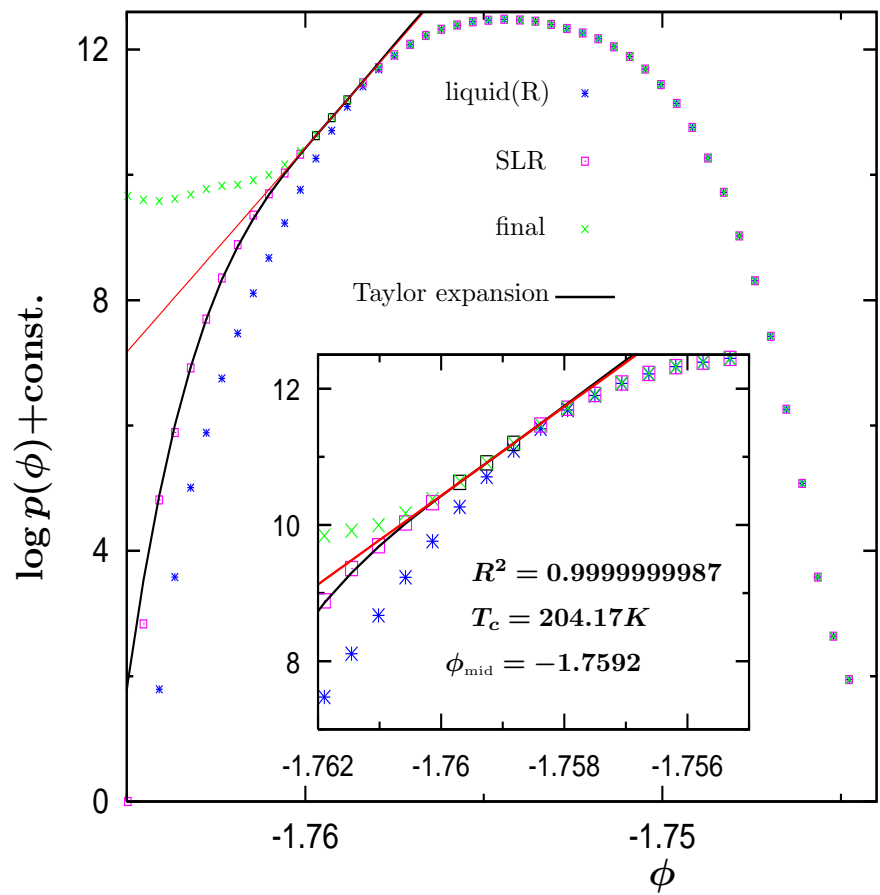
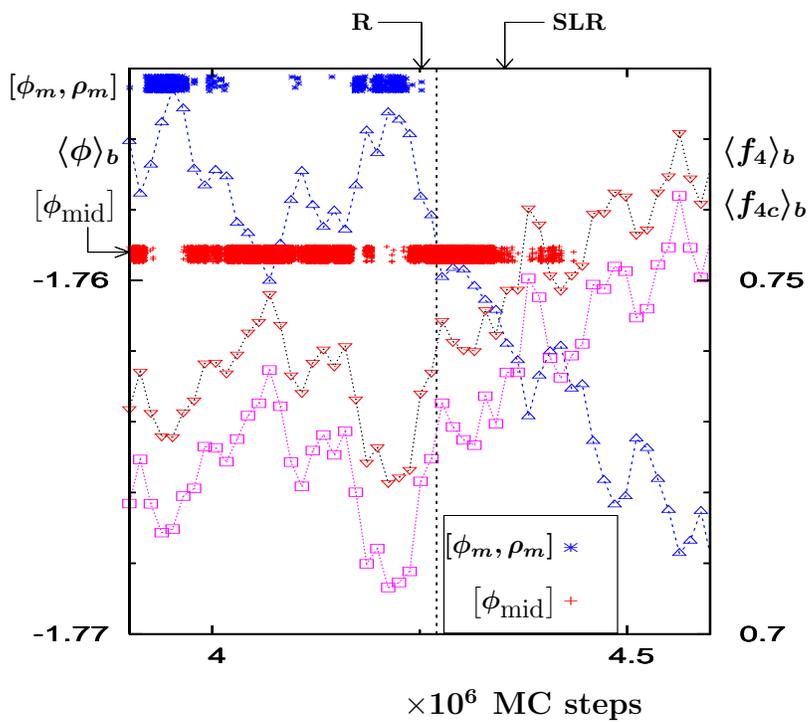
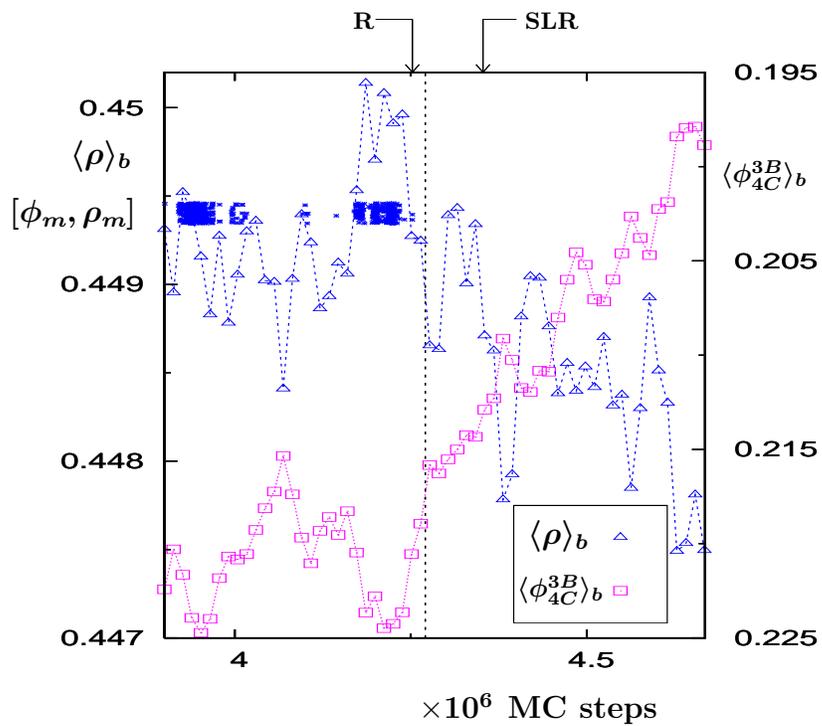
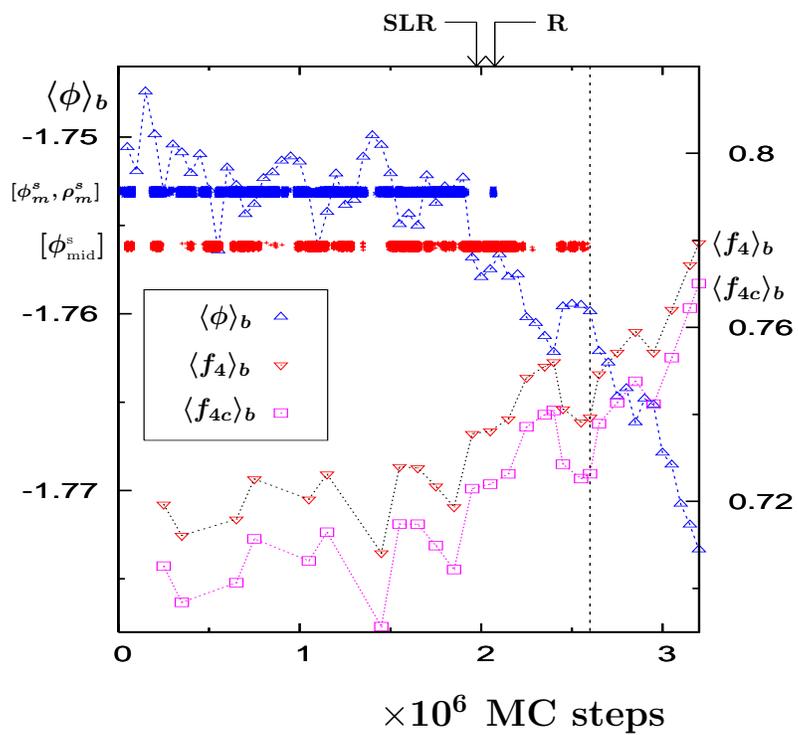
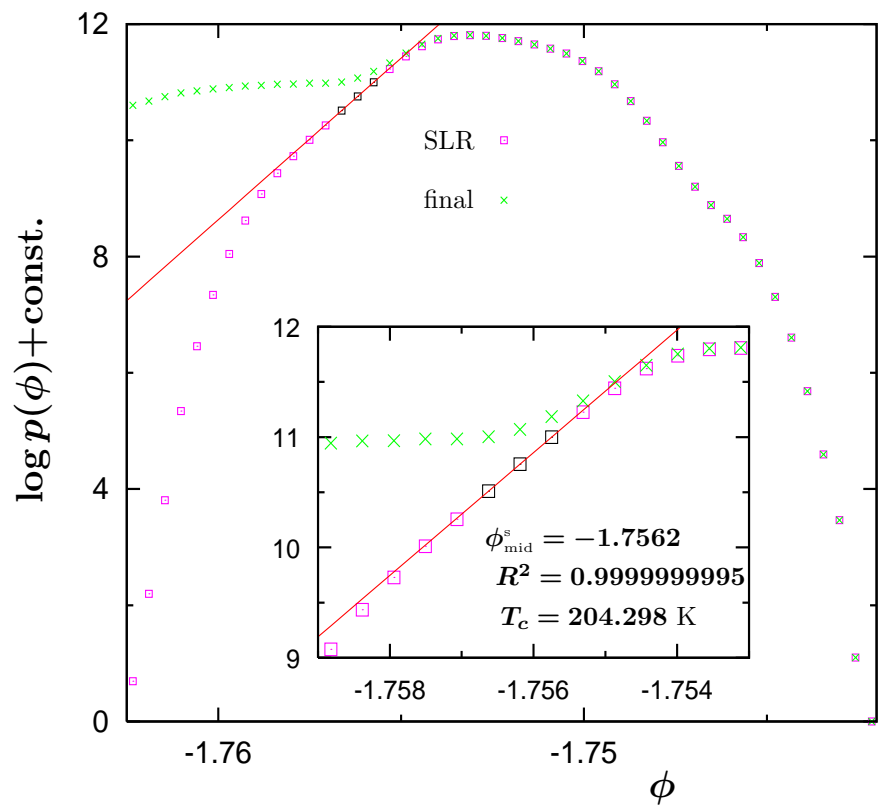
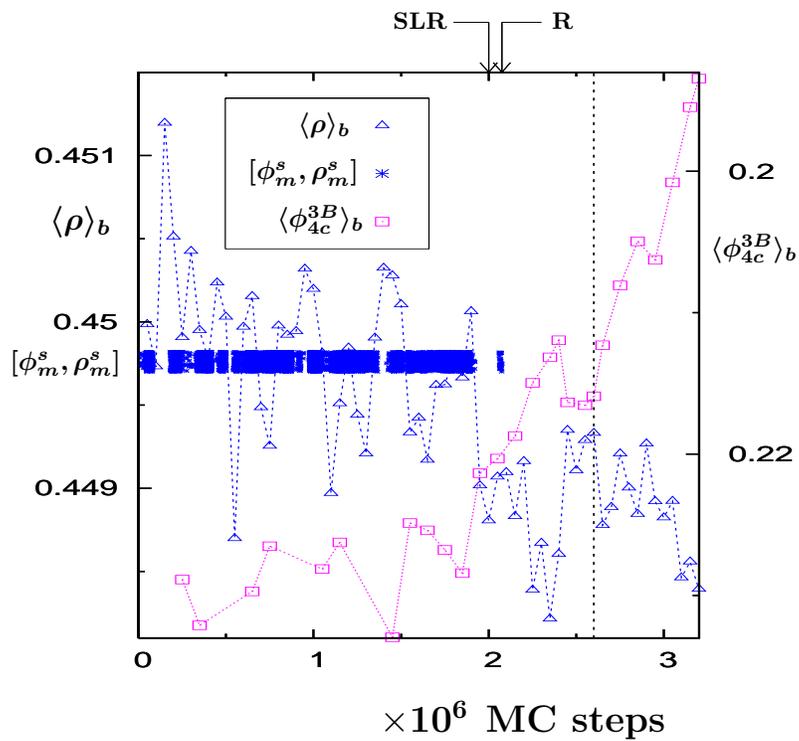
\end{document}